# Using Dust Shed from Asteroids as Microsamples to Link Remote Measurements with Meteorite Classes

B. A. Cohen[1], J. R. Szalay[2], A. S. Rivkin[3], J. A. Richardson[1], R. E. Klima[3], C. M. Ernst[3], N. L. Chabot[3], Z. Sternovsky[4,5], and M. Horányi[4,6]

[1]NASA Goddard Space Flight Center, Greenbelt MD 20771 (Barbara.A.Cohen@nasa.gov)
[2]Princeton University, Princeton, NJ 08544
[3]Applied Physics Laboratory, Johns Hopkins University, Laurel MD 20723
[4]LASP, University of Colorado, Boulder CO 80303
[5]Smead Aerospace Sciences, University of Colorado, Boulder CO 80309
[6]Physics Department, University of Colorado, Boulder CO 80309

**ABSTRACT**: Given the compositional diversity of asteroids, and their distribution in space, it is impossible to consider returning samples from each one to establish their origin. However, the velocity and molecular composition of primary minerals, hydrated silicates, and organic materials can be determined by *in situ* dust detector instruments. Such instruments could sample the cloud of micrometer-scale particles shed by asteroids to provide direct links to known meteorite groups without returning the samples to terrestrial laboratories. We extend models of the measured lunar dust cloud from LADEE to show that the abundance of detectable impact-generated microsamples around asteroids is a function of the parent body radius, heliocentric distance, flyby distance, and speed. We use monte carlo modeling to show that several tens to hundreds of particles, if randomly ejected and detected during a flyby, would be a sufficient number to classify the parent body as an ordinary chondrite, basaltic achondrite, or other class of meteorite. Encountering and measuring microsamples shed from near-earth and main-belt asteroids, coupled with complementary imaging and multispectral measurements, could accomplish a thorough characterization of small, airless bodies.

# 1. INTRODUCTION

By number, asteroids dominate the inventory of the inner solar system. Over half a million asteroids are known, and tens of thousands more are discovered each year. The asteroid population has members ranging from unaltered, primordial objects to remnants of differentiated parent bodies. Despite being composed mostly of silicate minerals, asteroids may have been important vectors for the delivery of water and organic materials to the early Earth (Morbidelli et al., 2000); recent measurements by the Rosetta spacecraft are inconsistent with a major cometary contribution (Altwegg et al., 2014) leaving asteroids as the likeliest exogenic contributor. Water or $OH^-$, bound into clay minerals, is present in amounts up to 10% or more by mass in carbonaceous chondrite meteorites (Brearley, 2006) and many asteroid classes (Rivkin et al., 2013; Rivkin et al., 2002; Volquardsen et al., 2007). Organic material has been detected both in meteorites and main-belt asteroids (Campins et al., 2010; Pizzarello et al., 2006; Rivkin and Emery, 2010). Thus, asteroids are important targets for both scientific inquiry, addressing the NASA goal of exploring and observing the objects in the solar system to understand how they formed and evolve, and also for sustaining and expanding destinations for human exploration and resource utilization.

Telescopic observations are the primary tool for assessing the range of asteroid compositions and linking them to meteorites (Table 1) (Burbine and Binzel, 2002; Bus and Binzel, 2002; Reddy et al., 2015; Xu et al., 1995). The majority of these measurements focus on the 0.5-1.0 µm spectral range, which provides identification of major iron-bearing silicates, but only a small fraction of observations extends to the 2.5 µm region to capture robust measurements of a range of silicate, carbonate, and accessory mineral spectral features (Figure 1). Focused observations integrating multiple wavelength regions are able to more effectively identify specific mineralogies, but the Earth's atmosphere severely limits many of these observations, including those of water and carbon-bearing materials. Compositions and meteorite linkages derived from telescopic spectra have been validated by spacecraft data to date for Eros, Itokawa, and Vesta (Binzel et al., 2001; Nakamura et al., 2011; Reddy et al., 2013; Fujiwara et al., 2006; Veverka et al., 2000).

More definitive links between meteorites and asteroids may be made by analyzing samples of the parent body directly. The Hayabusa mission recovered thousands of sub-100 µm particles, linking the asteroid Itokawa to LL chondrites (Nakamura et al., 2011). The Hayabusa 2 and OSIRIS-REx missions will collect more significant masses of regolith from asteroids to return to Earth for analysis (JAXA, 2016; Lauretta et al., 2015). Sample return missions are costly and inevitably limited to a small number of asteroids. However, in-flight composition analysis of regolith microsamples shed from asteroid surfaces may present good alternatives to direct surface access by a lander (Kempf, 2009; Postberg et al., 2011; Krüger et al., 2003; Mann and Jessberger, 2010; Postberg et al., 2014; Szalay and Horányi, 2016c). In this paper, we investigate how well measurements made by dust analyzer instruments map to the chemical characteristics of meteorites, conduct statistical exercises to determine the number of particles required to tie such measurements to known meteorite groups, and model the dust cloud density around near-earth and main-belt asteroids to predict whether a sufficient number of particles could be encountered by a spacecraft-borne dust analyzer during asteroid flyby missions to confidently make this measurement.

## 2. IN SITU DUST PARTICLE ANALYSIS

Compositional analysis of dust particles encountered by spacecraft has been successfully conducted using dust detector instruments coupled with an impact-ionization mass spectrometer. Planetary, interplanetary, and interstellar dust grains have been analyzed *in situ* by dust instruments onboard spacecraft such as Vega-1 and Vega-2 (Dikov et al., 1989; Kissel and Krueger, 1987), Giotto (Kissel et al., 1986; Kissel, 1986), Helios 1 (Altobelli et al., 2006; Grün et al., 1980), Stardust (Kissel et al., 2003; Kissel et al., 2004) and Cassini (Srama et al., 2004, Altobelli et al., 2016). Individual particles' origin as planetary or interplanetary is determined from its impact speed and direction.

The interpretation of impact mass spectra in terms of elemental ratios and mineralogy is aided by laboratory calibration measurements and remotely sensed data. Dust samples from relevant minerals or homogenous synthetic materials can be prepared and accelerated to typical flyby velocities using electrostatic dust accelerators, enabling laboratory calibration measurements that can be used to quantitatively interpret the mass spectra; corresponding mineralogy can be inferred from elemental ratios (Altobelli et al., 2016; Fiege et al., 2014; Goldsworthy et al., 2003; Hillier et al., 2014; Hillier et al., 2018). Many organic compounds can also be identified and quantified even at trace abundances (~1 ppm) (Kempf et al., 2012; Postberg et al., 2018).

As an example of the utility of *in situ* particle analysis and its fidelity, take the case of measurements of the composition of cometary dust. Cosmic dust samples have been extensively analyzed in the laboratory (e.g., Brownlee, 1985) to provide reference compositions for these particles, so these studies illustrate the parameters and precision needed to make the links between in situ and lab measurements. Jessberger (1999) summarizes the results from the PUMA-1 instrument on Vega-1 sent to Comet P/Halley, and Sekanina et al. (2001) summarize the results from this detector as well as those from PIA on Giotto and PUMA-2 on Vega-2; in total these detectors measured more than 2500 individual grains using a mass spectrometer with mass resolution ~100. At the high velocity of these dust impacts (typically above 20 km/s), the grains were completely vaporized and yielded cation spectra easily interpreted as elemental (rather than molecular) ions. Using the elemental ratios of Mg, Fe, Si, C and S, (Fomenkova et al., 1992) classified 2000 grains as metals, sulfides, silicates, and carbonaceous (CHON) particles, and identified organic compounds, pure-C grains, iron-rich oxides and carbonates. The mineralogical composition of Halley dust was estimated to be > 20% Mg-silicates, $\approx$ 10% Fe-sulfides, 1-2% Fe metal and <1% Fe oxide, leading to the conclusion that most of the comet Halley dust particles have an overall chondritic composition along with minor phases inferred to have a secondary origin.

Another successful example of effective and informative *in situ* grain characterization comes from the analysis of interplanetary, interstellar, and planetary dust particles by the Cassini Cosmic Dust Analyzer (CDA). CDA consisted of a dust detector and time-of-flight impact mass spectrometer, with a mass resolution of 10-50 over a mass range of 1-190 amu (Postberg et al., 2006). CDA returned two mass spectra of particles ~2-6 nm in size during its cruise phase to Jupiter, which consisted primarily of Fe, with possible Ni and Cr, indicating they were composed of iron or kamacite (Hillier et al., 2007). Cassini encountered and characterized a further 36 grains classified as interstellar dust particles based on their high entry speed and direction. Their major elemental ratios (Mg/Si, Fe/Si, Mg/Fe, and Ca/Fe) are interpreted as representing magnesium-rich grains of silicates, oxides, and Fe-Ni metal in ratios consistent with CI chondrites (Altobelli et al., 2016). At Saturn, the CDA recorded thousands of mass spectra of E-ring particles derived from Saturn's moon Enceladus. These particles have radii of 0.1-1 mm and are composed of water ice, water ice with significant organic and/or siliceous material, and water ice particles with particularly high sodium contents, providing evidence for a subsurface ocean on Enceladus in contact with a silicate seafloor (Postberg et al., 2009). CDA also observed silica particles <10 nm in size that were originally embedded in the icy particles, products of alkaline hydrothermal activity at the bottom of such an ocean (Hsu et al., 2015).

## 3. LINKING INDIVIDUAL PARTICLE COMPOSITION TO METEORITES

The examples of grain analysis from cometary tails, interstellar dust, and E-ring particles are examples of multiple detections of particles from a relatively well-known and homogeneous source. Asteroids, and their meteorite building blocks, are diverse in their chemical and mineralogic makeup, reflecting the range of parent bodies and geologic evolution of the solar system. In this section, we consider whether compositional information obtained on individual particles shed from asteroids can effectively to link them to known meteorite groups. We consider a simple case, where an asteroid's surface is made of a single meteorite type, and is shedding unaltered, monomineralic particles for detection by a flyby instrument. In the discussion, we will consider adjustments to the simple case based on observations from current spacecraft.

In the laboratory, meteorites are classified using petrologic characteristics (e.g., texture, mineralogy), elemental composition, and isotopic composition, and classification ranges from iron and nickel metals from protoplanetary cores to primitive stony meteorites that have never been significantly heated; along with some meteorites that do not fit in existing groups. While these measurements are more complex than can be expected for dust analyzers, previous missions showed that estimates of major mineral abundance (silicates, oxides, sulfides, metal), and elemental ratios could be obtained at a level sufficient to classify particles as IDPs. This same combination of mineral abundance and mineral composition is also generally sufficient to classify a bulk sample within the major meteorite groups; that is, as an ordinary (H, L, LL) chondrite, carbonaceous (C) chondrite, enstatite (E) chondrite, Rumuruti (R) chondrite, basaltic achondrite (eucrite, diogenite, angrite, aubrite, ureilite), primitive achondrite (brachinite, acapulcoite, lodranite, winonaite), or metal-rich achondrite (mesosiderite, pallasite, iron) (Rubin, 1997). To illustrate this, we used compilations of published point-counts to compile the average abundance of silicates, oxides, sulfides, metal, and the Fe/Mg ratio in silicates in meteorite groups (Table 1) (Bischoff et al., 2000; Bland et al., 2010; Bowman et al., 1996; Buseck, 1977; Jambon et al., 2005; Kallemeyn et al., 1996; Keil, 1962; Keil, 1968, 2012; Mayne et al., 2009;

Mittlefehldt et al., 1998; Rubin, 2007; Singletary and Grove, 2003; Weisberg et al., 2006).

Figure 2 graphically shows two possible combinations of parameters, the fraction of silicate particles vs. the Fe content in the silicates, and ternary composition of abundance of Fe-Ni metal, oxides, and sulfides+phosphates. Figure 3 provides another representation that simultaneously considers all parameters using Agglomerative Hierarchical Clustering (AHC), an iterative classification method where each meteorite class is compared with each other class, and successively merged (or agglomerated) into clusters until all classes have been merged into a single cluster that contains all classes (AHC was performed using the XLStat package for Excel, selecting the Pearson correlation coefficient for similarity and unweighted pair-group average for aggregation). In general, Figures 2 and 3 show that these parameters are sufficient to distinguish major groups of meteorites from each other, but also highlights where potential confusion may arise, requiring additional information to help distinguish the classes. However, it is highly unlikely that analysis of such microsamples would be the only aim of a space mission. Contextual clues provided by other observations of the parent body (for example, imaging and multispectral/hyperspectral analysis), would initially narrow the possible interpretation space. For example, pallasites and enstatite chondrites are similar to each other using only these parameters, whereas they would be easily distinguished in infrared spectroscopy. Additionally, several types of meteorites may be recognizable by the presence of unique minerals, for example: oldhamite ($(Ca,Mg)S$) in enstatite chondrites, graphite (C) in ureilites, hibonite and mellilite ($CaAl_{12}O_{19}$; $Ca_2MgSi_2O_7$) from calcium-aluminum inclusions (CAIs) in chondrites, cohenite ($(Fe,Ni)_3C$) in iron meteorites, and organic molecules and serpentines ($(Mg,Fe)_6Si_4O_{10}(OH)_8$) in carbonaceous chondrites. Other meteorite groups have distinctive elemental ratios in their silicate fraction, for example, calcium-rich pyroxene (fassaite) coupled with high-Ca plagioclase (anorthite) in angrites, and ordinary chondrites have very small ranges of Fe/Mg in silicates whereas the rare R chondrites exhibit a wide range of Fe/Mg ratios.

In this exercise we consider a minimum scenario, which is how well classification based solely on the abundance of major phases (silicates, Fe-Ni metal, sulfides, phosphates, and oxides) and the Fe/Mg ratio (or Fe content) of the bulk silicate component could link microsamples shed from an asteroid to a unique meteorite class. We evaluated the number of particles that would need to be analyzed to accurately link it to a specific class of meteorites using a combination of a Monte Carlo method of generating sample sets and a multidimensional nearest-neighbor matching algorithm. We generated sample sets of *n* particles from the four mineral types (silicate, sulfides+phosphates, oxides, and Fe/Ni metal) by choosing random numbers from a uniform distribution [0,1] and assigning each number a particle type depending on its abundance in Table 1 (e.g., values between 0-0.692 would be labeled "silicate" in an enstatite chondrite that has an expected silicate particle abundance of 69.2%). We then compared the particle abundances to each meteorite class by calculating the probability that a sample from each class would create each random sample. The conditional probability of a specific abundance of particles of different types, $a_1$, …, $a_n$, comprising a sample, *s*, given the sample matches a meteorite of class *b*, is calculated as

$$P(s|b) = \left(P(a_1|b)^{s(a_1)} \cdot , ... , \cdot P(a_n|b)^{s(a_n)}\right)^{-1}$$

where $P(a_1|b)$ is the probability of selecting particle of "type 1" once from the candidate class, and $s(a_1)$ is the number of "type 1" particles in the random sample.

We calculated the most likely meteorite class match to each simulated sample by evaluating the relative likelihoods of each candidate class yielding the sample. The relative likelihood of a class yielding a given sample is:

$$LR_b = \frac{P(s|b)}{P(s|\neg b)}$$

where $P(s|\neg b)$ is the probability of the sample not being created from the candidate class *b*. Assuming that the random sample actually does belong to one of the meteorite types in Table 1, this probability is the sum of probabilities $P(s|b)$ calculated for all classes except the class in question. In other words, $\sum P(s|\neg b_i)$ = unity for all meteorite classes, $b_1$, …, $b_n$. The likelihood ratio, LR, is calculated for each class and the sample is assigned to the most likely class. The evidence strength of the selected class was evaluated using log-scale thresholds given by Jeffreys (1998) (Table 2). The basic assumption in evaluating the strength of the evidence that links a random sample to a meteorite class is that the confidence in the selection is proportional to the likelihood ratio corresponding to the selected class. As the likelihood ratio of the selected class increases, it provides stronger evidence that the sample truly belongs to the meteorite class.

We then constructed numerical test datasets derived from individual meteorite characteristics and applied this approach to see whether our method would correctly determine their class. We created numerical test datasets containing the published mineral abundance and Fe/Mg ratio in silicates of Bluff (L5), Bath (H4), Abee (E), Juvinas (eucrite), ALH A77801 and Dhofar 1222 (acapulcoites). (Table 3). Each test dataset contained 1100 particles and each particle had an identity proportional to the abundance in Table 3. From each test dataset, we conducted interval trials, where each trial selected $n$ particles from the probability distribution to create a "sample" that we assumed to be the bulk composition of the meteorite. We then compared the bulk composition derived from the $n$ particles in the test dataset to the composition of the 14 meteorite classes in Table 1 to determine the best-fit class for each "sample." We varied $n$ from 10 to 1100 in intervals of 20 and conducted 300 unique trials at each interval $n$. This approach provides the ability to probe the confidence in a meteorite class assignment as a function of the number of particles in a sample dataset.

The results are shown Figure 4. For Bluff and Bath, only about 100 particles were needed for 90% of trials to determine they are in the ordinary chondrite class; the specific subclass (L and H, respectively) became apparent in samples sizes of about 200 particles. The Abee enstatite chondrite and Juvinas eucrite were easily classified even with a few tens of particles. The model also illustrates where the method of using only average compositions may fail; for example, both acapulcoites were misclassified but not consistently (L chondrite and enstatite chondrite). However, it should be reiterated that additional information (multispectral data, range of composition within silicate minerals, etc.) would be likely be available during a mission to help distinguish these (and other) meteorite classes.

We applied our approach to the Hayabusa sample data to investigate whether a similar approach would have been able to link Itokawa samples to a specific meteorite group, if they had been encountered *in situ* rather than returned to the Earth. One consistent database of 1087 monomineralic Hayabusa particles shows that 83% are silicates (olivine, pyroxene, and feldspar) and these silicates have an Fe/Mg ratio of 0.43 (Nakamura et al., 2011) (Table 1). We used these abundances to generate numerical test datasets and trials as above. The best-fit parent body for each trial using the Hayabusa dataset is illustrated in Figure 5. For trials of smaller numbers of particles (<100), the samples may appear to derive from seven of the 14 parent bodies considered, having some chance of being similar to basaltic achondrites, ordinary chondrites, carbonaceous and enstatite chondrites. At around 100 particles, the sample is clearly distinguished (90% of all trials) as an ordinary chondrite, though multiple subtypes are still in play. When *n* reaches 200 particles, the sample is consistently closest to an LL-type ordinary chondrite (90% of all trials). However, only 60% of the *n*=200 assignments are made with evidence that can be considered "substantial" or stronger. The evidence that the meteorite class is LL-type and not another class is "strong" or higher at *n*=700. In trials with large numbers of particles, the second-choice class is the very similar L chondrites, where the occurrence of weakly-evidenced selections is proportional to the occurrence of L-type selections. When the sample size reaches the actual number of Hayabusa returned particles (1087), the selected class is LL-type chondrites with an LR of 1020, which is decisive evidence that the sample is derived from an LL-chondrite. In the returned Hayabusa samples, their origin as LL chondrites was unequivocally established using oxygen isotopes and trace-element analyses (Nakamura et al., 2011).

The Stardust mission returned particulate samples from the coma of comet Wild 2. The elemental composition of the Stardust samples are in broad agreement with the analyses made by the dust analyzer aboard the Halley mission (Flynn et al., 2006 ). Most of the particles are weakly constructed mixtures of nanometer-scale grains with occasional much larger (>1 mm) ferromagnesian silicates, Fe-Ni sulfides, and Fe-Ni metal (Zolensky et al., 2006). Organic molecules are also found in many of the grains (Sandford et al., 2006 ). However, many more monomineralic fragments were created by the breakup of the cometary particles upon capture in the aerogel (Joswiak et al., 2012). The characteristics of these <10 μm monomineralic particles, when examined as a microsample set, do not appear to link Wild 2 to carbonaceous chondrites, as might be expected, given the similarity of IDPs to CI meteorites (Figure 2). It may be that the Stardust dataset is too small to yield reliable statistics (*n*=95), or that the bulk composition of Comet Wild-2 is not well represented by the subset of monomineralic grains. However, the Stardust samples are linked to IDPs by methods available in the laboratory, including trace elements in olivine, amorphous silicates known as GEMS, and the morphologic presentation of pyroxene as platelets and whiskers (Brownlee et al., 2006).

## 3. DUST CLOUD GENERATION AROUND ASTEROIDS

Having shown that mass spectroscopy of grains can be used to link microsamples to meteorite groups, we now consider the number and density of particles shed from asteroid parent bodies available to make such analyses. With the exception of active asteroids, a key process for such shedding is meteoroid impacts, which eject material, both neutral and charged, that become part of the environment, either as bound or unbound particles, around all airless bodies (Szalay and Horanyi, 2016a). The tenuous dust cloud is maintained by the continual bombardment of the surfaces by fast, interplanetary micrometeoroids. Such ejecta clouds were detected and characterized *in situ* by the Galileo mission during flybys of the icy moons of Jupiter (Europa, Ganymede, and Callisto), and also around the Moon by the LADEE mission (Horányi et al., 2015; Krivov et al., 2003; Krüger et al., 1999; Krüger et al., 2000; Sremčević et al., 2005; Sremčević et al., 2003; Szalay and Horányi, 2015b).

To model micrometeorite-derived ejecta clouds near asteroids, we build upon an existing model for the dust distribution around asteroids near 1 au (Szalay and Horányi, 2016a). This model employs the same initial velocity distribution of ejecta particles as inferred from measurements by the Lunar Dust Experiment (LDEX) on LADEE and assumes that the same distribution holds at other airless bodies near 1 au. LDEX was not a mass analyzer, but rather a very sensitive impact ionization dust detector capable of individually detecting dust grains with radii a > 300 nm (Horányi et al., 2014). Throughout its operation from orbit, LDEX discovered that the permanently present, asymmetric dust cloud at the Moon is sustained primarily by the helion (HE), apex (AP), and anti-helion (AH) sporadic meteoroid sources, with the AP being the strongest producer of ejecta (Szalay and Horányi, 2015a). A minor contribution from the anti-apex (AA) source near dusk was also noted in subsequent analysis to account for the low but appreciable densities near dusk (Szalay and Horányi, 2016c).

At the Moon, the vast majority of grains in the ejecta cloud are gravitationally bound (Horanyi et al., 2015). On the other hand, for the much smaller asteroids, essentially all grains are unbound; bound grains only make up a few percent of the ejecta cloud even for 100 km radius airless bodies (Szalay and Horányi, 2016a). Hence, with the exception of Ceres and Vesta, we assume all asteroids from 1-3 au have negligible gravity with respect to the dynamics of ejected grains. With this assumption in place, the dust density distribution in the ecliptic plane near a small airless body is

$$n(r, \varphi, a, d) = n_w \left(\frac{R}{r}\right) a_\mu^{-2.7} \sum_s w_s(d) \cos^3(\varphi - \varphi_0) \Theta\left(|\varphi - \varphi_0| - \pi/2\right) \quad \text{(Eq. 1)}$$

where $n_w = 5 \times 10^{-4}$ m$^{-3}$, $R$ is the radius of the airless body, $r$ is the distance from the center of the airless body, $a_μ$ is the particle radius in μm, $s$ is the index indicating the source (HE, AP, AH, or AA), $w_s$ is the relative weight of the source that is a function of the heliocentric distance $d$, $φ_s$ is the angle of the given source from apex, and Θ is the Heaviside function.

To extend this model to different heliocentric distances, we must make some additional assumptions about the sporadic sources. The HE/AH sources are generated by prograde short period Jupiter Family Comets (JFCs) (Nesvorny et al., 2011a). The AP source is generated by particles on retrograde orbits shed from both Oort Cloud Comets (OCCs) and Halley Type Comets (HTCs) (Nesvorny et al., 2011; Wiegert et al., 2009). While the retrograde OCC and HTC grains are significantly less numerous than JFCs at 1 au, the measured lunar dust cloud was found to be primarily caused by these particles as their impact velocities are so large. Quantitatively, the production of impact ejecta is given by

$$w_s = C\, F_{imp}(d) m_{imp}^{α+1}(d) v_{imp}^{δ}(d) \qquad \text{(Eq. 2)}$$

where $C$ is a fitting constant, $F_{imp}$ is the incident number flux, $m_{imp}$ is the characteristic mass, $v_{imp}$ is the incident velocity, and $α = 0.2$ & $δ = 2.5$ are fit constants from laboratory impact ejecta experiments (Koschny and Grün, 2001; Krivov et al., 2003). At 1 au, $w_s(1)$ = [0.24, 0.49, 0.24, 0.03], building on previous analyses (Szalay and Horányi, 2016a; Szalay and Horányi, 2016c).

To determine the production rate, $w_s(d)$, for heliocentric distance $d > 1$ au, we use Eq. 2 and make four further assumptions: a) impactor fluxes from the relative sources scale as given in Poppe (2016); b) characteristic masses do not change as a function of $d$; and c) that all asteroids are on circular Keplerian orbits and all impact velocities scale like $1/\sqrt{d}$, following Kepler's law. For the flux scaling, we use the values given in Poppe (2016) for impactor fluxes onto a body in a circular orbit for various sources as a function of $d$ for 100 μm grains. Since the OCC grain flux decreases faster than HTCs with distance from 1-3 au, we assume the AP source scales as HTCs. For the HE/AH sources, we use the scaling from the JFCs in this model. The assumption that the impactor velocities scale as $1/\sqrt{d}$ is certainly reasonable for the AP source, as to first order this can be approximated as head on collisions between circular retrograde grains and the circular prograde orbit of the airless body. The AA source is relatively unimportant in this scenario, scaling approximately following Kepler's law. For the HE/AH velocity scaling, without a complex particle tracing model to predict the change in impacting velocity distribution function, we also use Kepler's law to estimate a reasonable zeroth-order velocity scaling. With these assumptions and utilizing the values of $w_s(d=1)$ as previously determined (Szalay and Horányi, 2016a), we can then numerically calculate $w_s(d)$. A targeted study on eccentric asteroid 3200 Phaethon revealed that the eccentricity or the asteroid can modulate the expected ejecta production (Szalay et al., 2019a), however, we fix the orbits to be circular in this study to provide a baseline comparison. One last assumption relating to Eq. 1 is that the relative angle from apex for the HE/AH sources remains the same from 1 to 3 au. While this angle would most likely also change to a certain extent as a function of $d$, without significant additional modeling to inform on this angle, we make the simplifying assumption that it remains constant.

Figure 6 shows the calculated ejecta cloud density distribution for a 10-km radius airless body at 1 and 3 au, for grains with radii $a > 50$ nm. The lily pad structure of the ejecta cloud at 1 au is identical to previous work (Szalay and Horányi, 2016a), and shown in a reference frame with the Sun in the -x direction and apex (the direction of orbital motion) in the +y direction. The impact rates at dusk remain poorly constrained, hence the density of the ejecta cloud over this hemisphere remains only a qualitative estimate.

The most apparent and expected trend exhibited in Figure 6 is the decrease in absolute number density as a function of heliocentric distance. Since both the fluxes and impact velocities decrease with heliocentric distance, the subsequent impact ejecta production also decreases following Eq. 2. A secondary effect is the slight modification of the ejecta cloud structure due to changes in the relative importance of the various sources. Because the velocities are all forced to follow the same scaling and the relative angles remain the same, a consequence of the assumptions outlined above, the primary modification in structure arises from relative differences in flux between the AP and HE/AH sources. Since the HTC flux decreases more rapidly than the JFC flux, the AP source is diminished at 3 au on the right panel of Figure 6. This flattened structure is a direct consequence of the model used and may vary for different cometary dust models. Overall, however, the structure is similar for ejecta clouds around asteroids from 1 to 3 au, namely that the density distribution is always enhanced on the apex side and the density will decrease for increased heliocentric distance.

In order to predict impact rates and the total number of particles a dust detector would encounter on a flyby mission to an asteroid, we first investigate the minimum dust size for a reliable detection and analysis. Assuming an LDEX type instrument with a similar sensitivity of 3,000 electrons of impact plasma necessary to individually detect an impacting dust grain and an impact plasma dependence of $Q(v) \propto mv^\beta$, we calculate the minimum detectable size as a function of impact velocity. Due to the low impact velocities (about 1.6 km/s), the LDEX instrument was able to verify the power law character of the lunar ejecta dust distribution only for ≥300 nm sized particles, with a cumulative particle size distribution index of -2.7. As an approximation, we assume that the size distribution function of ejecta particles from asteroids follows a similar law down to 50 nm. To calculate total impact numbers, we assume the effective detector area of 600 cm$^2$, which is the capability of recently developed dust analyzer instruments.

Figure 7 shows the minimum detectable particle size for four separate calibration curves. The original calibration of LDEX reported an impact velocity dependence of $\beta = 4.74$ (Horányi et al., 2014). More recent calibrations indicate the exponent may be $\beta = 5.0$, therefore we calculated the size dependence for both exponents. Additionally, we use two different mass density values, 2.5 g/cc and 3.33 g/cc to represent a reasonable range for asteroidal ejecta particles. The subsequent four curves then represent an approximate bounding parameter space for the size dependence, which is similar for all curves. As an example, impact (or flyby) velocities of ~5 km/s would allow for individual detection of grains with $a$ ≥50 nm. We use a detection threshold of $a$ ≥50 nm for the following analyses.

With the size threshold in place, using Eq. 1 and the analysis discussed in the previous section, we can determine the dust density distribution near an asteroid and flyby impact rates. Figure 8 shows an example flyby for a body of radius $R$ = 10 km at heliocentric distance $d$ = 1.5 au. The ejecta cloud density distribution is a hybrid of the 1 and 3 au cases (Fig. 6). The left panel shows the encountered number density along the flyby (at various altitudes, $h$, above the surface) as a function of distance from closest approach, as well as the total cumulative impacts. The solid lines denote apex-hemisphere flybys. Note, the values of $h$ for each flyby are different from the impact parameter (measured from the center of the body) shown in a previous, similar analysis (Szalay and Horányi, 2016a). The left panel shows the number density along the flyby as a function of distance from closest approach as well as the total cumulative impacts. Transits on the apex side (solid lines) result a factor of ~3 times larger number of total impacts for a given flyby altitude. As evidenced by the impact profile, the majority of the impacts occur within ±100 km from closest approach.

Similar analyses can then be repeated over a grid of heliocentric distance and flyby altitude for various asteroid sizes to determine the optimum parameters for a required number of collected dust impacts. Figure 9 shows the total number of expected dust impacts as a function of d and flyby altitude for detection particle size threshold $a \geq 50$ nm. The results can be scaled to a larger dust size threshold, $a > a_0$, using a $N(a) = N_0(a/a_0)^{-2.7}$ where $N_0$ is the number of impacts given in Figure 9 and $a_0 = 50$ nm. The largest impact rates would be measured for flybys near the surface, at smaller heliocentric distances, and near larger bodies. The specific dependencies of the total impact count on body radius, altitude, and heliocentric distance are a function of both the geometry of the flyby orbit and the relative strengths of the various dust sources.

It should be noted that this analysis represents only a lower limit to the dust density. There are additional sources of dust that could modify the size/number distribution of ejecta, many of which may have relatively low ejection velocities but could contribute to the overall number density. LDEX was only able to determine the dust density distribution down to altitudes of approximately 1 km, which corresponds to particles which left the surface with approximately 60 m/s. It is possible that the velocity distribution function has an enhancement at initial speeds lower than this. However, for the 50 nm grains discussed primarily in this work, radiation pressure (RP) can significantly alter trajectories within 1000 km of the body for initial speeds on the order of 10's m/s, which could effectively sweep out such grains from the measured density distribution. Figure 10 shows an example of such a process. Here, we include only the force of RP, assuming the grains are spherical and perfectly reflecting and that the asteroid is at 1 au to represent the strongest case of RP. Dots show the location of particles launched from a uniform grid in angle from the surface of the airless body. As shown in this figure, particles with initial speeds on the order of 10 m/s or lower would experience a deviation from straight-line trajectories, modifying the dust density distribution.

## 4. DISCUSSION

*Linking particles to parent bodies.* Despite the wide diversity of meteorites and the complexities of their compositions, major mineralogy and silicate Fe/Mg ratios are generally sufficient to distinguish major groups of meteorites from each other, given a sufficient number of particles. The number of particles needed to make a decisive determination of the meteorite group varies but is generally in the hundreds of grains. This analysis assumes that the elemental analysis of each microsample is both precise and accurate. The precision and accuracy of the individual analyses depends on many factors, including the particle speed, ionization potential, instrument calibration, etc. However, because we have simplified the analysis to basic mineralogy, the measurement precision of each analysis need only be sufficient to recognize it as a silicate, metal, oxide, or sulfide, which is well within the capabilities of current instrumentation. Our approach also assumes that each grain is monomineralic, and that information about individual minerals can be extracted from the mass spectrum. The micrometer-sized grains in the Hayabusa and Stardust collections were usually polymineralic, but the proportion of monomineralic grains and fragments increased with smaller grain sizes, increasing the chances that a dust grain measuring 50-100 nm would be monomineralic. Even these small particles, when encountered at flyby speeds, would yield adequate elemental analyses, as they consist of tens of millions of atoms. Grains measuring a few nm and larger were successfully analyzed using the Cassini CDA (Hsu et al., 2015).

Our simple approach using only a few minerals and element ratios highlights where potential confusion may arise, requiring additional information to help distinguish the classes. Several types of meteorites may be recognizable by the presence of unique minerals that would be identifiable using additional elemental lines and ratios, for example: oldhamite ($(Ca,Mg)S$) in enstatite chondrites, graphite (C) in ureilites, hibonite and mellilite ($CaAl_{12}O_{19}$; $Ca_2MgSi_2O_7$) from calcium-aluminum inclusions (CAIs) in chondrites, cohenite ($(Fe,Ni)_3C$) in iron meteorites, and organic molecules and serpentines ($(Mg,Fe)_6Si_4O_{10}(OH)_8$) in carbonaceous chondrites. Other meteorite groups have distinctive minerals that occur together in their silicate fraction, for example, calcium-rich pyroxene (fassaite) coupled with high-Ca plagioclase (anorthite) in angrites. The dispersion of compositions can also be a clue to meteorite origin, for example, the equilibrated ordinary chondrites have a very small range of Fe/Mg in silicates whereas the rare R chondrites exhibit a wide range of Fe/Mg ratios.

Important contextual measurements (for example, imaging and multispectral/hyperspectral analysis), can constrain possible matches by narrowing the possible interpretation space. For example, pallasites and enstatite chondrites are similar to each other using only these parameters, whereas they would be easily distinguished in infrared spectroscopy. Take for example the NEAR mission, where global measurements of Fe, K, and Th indicate that the elemental composition of Eros' surface is consistent with that of chondrites (Peplowski, 2016), but constraints from gamma-ray and near-infrared spectroscopy rule out carbonaceous chondrites (McCoy et al., 2001), Ca/Si ratios rule out enstatite chondrites (Lim and Nittler, 2009), and x-ray spectral data rule out H chondrites (Foley et al., 2006). Taken together, the datasets indicate that Eros' measured surface composition is best matched by the L and LL chondrites, which would provide strong constraints to interpretation of microsamples encountered around this asteroid. These measurements of the surface of Eros and other asteroids, along with the Cassini measurements of dust particles derived from interplanetary sources (Hillier et al.., 2007) also illustrate that the surface materials of airless bodies are not hopelessly altered by billions of years of micrometeorite impacts. The surfaces of such bodies undergo space weathering, comminution, and likely contributions from impacting bodies, but their surfaces are recognizable in the context of meteorites (e.g., McSween et al., 2013); microsamples shed from these surfaces by spallation and electrostatic lofting would not be expected to have been rendered unrecognizable in the process.

*__Development of dust clouds around asteroids.__* Our modeling provides a lower limit to the amount of dust lofted from asteroids, modeling solely the contribution from micrometeorite impact. It is possible that other mechanisms may contribute appreciably to the number density, including surface grain size, jetting (as observed on Bennu, reference forthcoming) and electrostatically lofting. Recently-visited asteroids Bennu and Ryugu have an overall larger grain size and more exposed hard surfaces than the Moon's regolith (Lauretta et al., 2019; Watanabe et al., 2019). Impact ejecta observations and comparison to impactor models suggest that the Moon's fine regolith surface may have a lower yield than an equivalent solid surface of the same material, where a larger fraction of impact energy may get partitioned to local heating of the regolith instead of into the kinetic energy of ballistic ejecta (Szalay et al., 2019b, Pokorny et al., 2019). Therefore, a harder asteroidal surface could produce significantly higher ejecta yields than the Moon's regolith surface, from which these predictions were derived. Additionally, asteroids with large eccentricity will experience enhanced ejecta production on their ram hemisphere (Szalay et al., 2019a) and a flyby near an eccentric asteroid near its ram hemisphere would boost total dust detections.

While LDEX searched for a population of small (100 nm) grains electrostatically lofted to high altitudes, it found no evidence of such a population within the detection limits (Szalay and Horányi, 2015b). Additional remote sensing efforts at the Moon have also only yielded upper limits to the total population of high-altitude lofted dust (Feldman et al., 2014; Glenar et al., 2014; Glenar et al., 2011). These observations do not preclude the possibility of small-scale electrostatic transport ejecting appreciable quantities of dust. Recent laboratory experiments indicate a large degree of small-scale mobilization occurs for dusty surfaces exposed to relatively simple photo or plasma conditions (Wang et al., 2016). However, any process which ejects particles with speeds on the order of tens of m/s or less would be subject to radiation pressure as discussed for the low velocity ejecta. Whether such processes could significantly augment the highly energetic impact ejecta process remains unclear.

Several key assumptions were introduced to simplify the mathematical and modelling complexity involved in created impact ejecta predictions. These assumptions may have appreciable effects that might need to be considered in future mission planning. One key assumption is that all encountered bodies are on circular, Keplerian orbits. This serves to simplify the ejecta cloud structure, as for bodies with circular orbits we would expect the ejecta cloud to be approximately symmetric about the apex direction. Additionally, it allows us to directly utilize the Poppe (2016) results for fluxes onto circular orbits. However, for bodies which have sufficiently large eccentricities, large seasonal variations would be expected as the relative velocity vector between the body and the sporadic sources varies throughout the year. Since the ejecta response function is highly velocity dependent, eccentric bodies would have asymmetric ejecta distributions. Were a spacecraft to encounter an eccentric body, it would measure enhanced dust impact rates on the hemisphere of the body with the largest relative velocity to the HE/AH source. A prime example of this phenomena occurs at Mercury, whose large eccentricity causes highly asymmetric distributions in impact vaporization (Pokorný et al., 2017). Additionally, the relative angles of the HE/AH sources will also vary with both eccentricity and heliocentric distance.

For the nominal flybys considered, we calculated predicted impact rates for transits in the ecliptic plane. Were a flyby to transit with a non-ecliptic geometry, the non-equatorial sporadic sources would need to be included into the model, namely the northern and southern toroidal sources which are understood to be generated by highly inclined HTCs (Pokorný et al., 2014). While the toroidal sources are significantly lower in flux, their large impact velocities may result in these sources playing a large role in generating impact ejecta in mid and especially high latitudes, similar to how the AP source is the dominant ejecta producer at the Moon due to it large velocity. Even in the ecliptic plane, the relative contributions of the retrograde HTCs and OCCs generating the AP source to the JFCs generating the HE/AH sources remain relatively unconstrained past 1 au. Future modeling efforts could shed light on this issue and would improve estimates for impact ejecta distributions throughout the solar system.

Lastly, there is always the possibility, while small, that a given flyby occurs during a meteoroid shower at the target asteroid. During the LADEE mission, LDEX recorded significant enhancements due to meteoroid stream activity, most notably during the Geminids which caused an increase of tenfold in the number of dense ejecta plumes detected compared to the sporadic background (Szalay and Horányi, 2016b; Szalay et al., 2018). The meteoroid environment, specifically the meteoroid streams, are particularly unconstrained outside of 1 au. Yet, if the environment past 1 au is similar to Earth's in meteoroid activity, asteroids could experience a handful of intense meteoroid streams each of which lasts few days that could liberate significantly larger quantities of impact ejecta than the nominal sporadic sources.

***Instrument improvements:*** Even with modest performance (mass resolution of $m/\Delta m \sim 30$-100), dust detector-mass analyzer instruments have contributed major scientific discoveries, as discussed above. Over the past decade, much more capable *in situ* dust analyzer instruments have been developed (Srama et al., 2006; Sternovsky et al., 2007; Sternovsky et al., 2011). These instruments combine large effective detector areas ($\geq 600$ cm$^2$) with time-of-flight, reflectron-type impact mass spectrometers. The SUrface Dust Analyser (SUDA), selected to fly on board the Europa Clipper mission to measure the composition of ballistic dust particles populating the thin exospheres around the Galilean moons, has an effective mass resolution $m/\Delta m$ of 200-250 over mass 1-250 (Kempf et al., 2014). The Interstellar Dust Experiment (IDEX), selected to fly on the Interstellar Mapping and Acceleration Probe (IMAP) mission, will provide the elemental composition, speed and mass distributions of interstellar dust particles, with an effective mass resolution $m/\Delta m$ of ~200 over mass 1-500 (McComas et al., 2018). Together with ongoing technologic development, continued laboratory testing and calibration are continuing to improve the analytical robustness and accuracy of *in situ* microsample analysis.

## 5. CONCLUSIONS

The major meteorite groups may be distinguished from one another using indicators such as mineral abundance and elemental ratios. Several hundreds of microsamples derived from the meteorite bulk composition, if randomly ejected and encountered, would readily accomplish the goal of linking mineral abundance and composition with known meteorite types. Our modeling shows that hundreds of samples shed from a parent body could be collected and analyzed at relevant speeds during flybys of individual asteroids, depending on the mission design, and that these numbers are sufficient to robustly link dust samples to known meteorite classes.

The abundance of microsamples around asteroids is a function of the parent body radius, heliocentric distance, and altitude above the surface. Encountering thousands of these particles shed from larger, main-belt asteroids would be accomplishable with spacecraft that comes within a body radius of the surface. Hundreds of microsamples may still be measured from near-Earth asteroids or small main-belt asteroids during a close encounter (either by the main spacecraft or secondary payloads, where the secondary could make a low pass and report data to the primary), or by increasing the effective area of the dust sampling instrument. If smaller numbers of microsamples were encountered, constraints on parent body types might still be usefully inferred by examining the dispersion of compositions, detecting diagnostic phases (e.g., hydrated silicates, carbonates, organic molecules, and silicate compositions unique to specific meteorite groups, such as oldhamite or fassaite), and combining microsample analysis with other remote methods of determining composition, such as spectroscopy.

Dust analyzer instrumentation is steadily advancing, taking advantage of technological advancements in electronics, materials, and mechanical designs. Future dust detector instruments may be expected to have large sensitive areas, low mass, and improved dynamical sensitivity for compositional analyses. The velocity of incoming particles provides additional constraints on the source of the detected particles (e.g., interplanetary versus interstellar). With a high-performance dust trajectory sensor, particle ballistic paths may be traced back to their point of origin (Postberg et al., 2011), relating the measured composition of the grain to parent-body geologic features. Ideal flyby speeds for such analyses would be between 3 and 7 km/s to retrieve both elemental and molecular composition of the particles, but even at speeds greater than ~15 km/s, full elemental analysis may be conducted.

Given the compositional diversity of asteroids, and their distribution in space, it is implausible to consider returning samples from each one to establish their origin. However, sample return to Earth is not the only method for asteroidal sample analysis. The impact ejecta clouds that are continually sustained around asteroids provide a rich and so far untapped resource, and can provide crucial insight into understanding the origin and evolution of airless bodies in the solar system. *In situ* dust analyzers may be enabling for future missions to obtain the elemental and mineralogical composition measurement of dust particles originating from airless bodies without returning the samples to terrestrial laboratories (Rivkin et al., 2016).

## 5. ACKNOWLEDGEMENTS

This work was supported by the Johns Hopkins University Applied Physics Laboratory and the NASA Solar System Exploration Virtual Institute (SSERVI) – in particular, B. Cohen received support through the CLASS (Center for Lunar and Asteroid Surface Science) node, and M. Horanyi and Z. Sternovsky acknowledge support through the IMPACT (Institute for Modeling Plasmas, Atmosphere and Cosmic Dust) node. JRS was supported by NASA's Lunar Data Analysis Program (LDAP), Grant 80NSSC17K0702. A. Rivkin acknowledges support from the NASA Near Earth Objects Observation program grant NNX14AL60G. Thoughtful reviews by *your name here* strengthened this manuscript. We use the NASA Astrophysical Data System Abstract Service.

**TABLES & FIGURES**

Table 1: Asteroid classes and linked meteorite groups and their mineralogical characteristics.

Table 2. Threshold values and adjectival ratings of likelihood ratio (LR).

Table 3. Test case meteorite phase abundances, most likely model classification, and number of particles $n$ in the sample required to yield the most likely class (and subclass) in 90% of trials. In general, tens to hundreds of particles are sufficient to classify a sample using only these parameters, but in some cases additional information would be useful.

Figure 1. Example of silicate, carbonate, organic, and accessory mineral spectral features in the near IR region.

Figure 2. Meteorite groups plotted as simple combinations of a) silicate abundance and Fe/Mg ratio in the silicate fraction (unequilibrated meteorites have large ranges in Fe/Mg ratios); and b) sulfides + phosphates, oxides, and FeNi metal.

Figure 3. Agglomerative Hierarchical Clustering of meteorite classes in Table 1 using only the average abundance of four mineral phases (silicate, sulfides + phosphates, oxides, and Fe-Ni metal) shows that in general, multidimensional fitting using these parameters readily distinguishes major groups of meteorites from each other.

Figure 4. Classification results for model datasets created from known meteorites. a) Bluff and b) Bluff are linked to ordinary chondrites in about 100 samples, and to their subclass in about 200 samples. C) Abee and d) Juvinas are shown to be an enstatite chondrite and a eucrite using only 10's of samples. However, metal-poor acapulcoites e) ALH 77011 and f) ALH 77011 are confused with other classes.

Figure 5. (top) Selected meteorite classes for individual trials (y axis) of sample sets with $n$ from 20-1100 (x axis) selected from the Hayabusa returned sample (Table 1). As n increases, the number of trials corresponding to chondrites and specifically LL-chondrites, increases. (Bottom) the evidential strength of the selected meteorite class for each individual trial, no matter the guessed type. For small numbers of particles, the evidence to suggest that the sample matches one meteorite class over another is generally weak, but increases with increasing n.

Figure 6. Comparison of the ejecta cloud structure for a 10-km radius body at 1 au and 3 au for grain radii a > 50 nm. The reference frame shown has the Sun fixed in the –x direction and the apex direction in the +y direction. Due to the expected larger decrease in retrograde HTC particles compared to JFCs producing the HE/AH ejecta, ejecta from the apex source is reduced at 3 au producing a flattened density distribution.

Figure 7. Minimum detectable dust grain radius as a function of impact velocity.

Figure 8. Total number of impacts and predicted dust density (a > 50 nm) for a 10-km radius body at 1.5 au. Contours are logarithmically spaced in units of 10-3 m-3.

Figure 9. Total impact count predictions for grain size a > 50 nm as a function of heliocentric distance and flyby altitude for R = 10, 30, 70, and 100 km.

Figure 10. Simple simulation showing the effects of radiation pressure on ejecta dynamics, with the dots representing positions of grains launched from the surface at different times. Left and right are given for initial speeds of 10 m/s and 100 m/s respectively.

Table 1: Asteroid classes and linked meteorite groups and their mineralogical characteristics.

| Class | Group | Asteroid match | Silicate Fe/Mg ratio | Silicate (%) | Sulfides & Phosphates (%) | Oxides (%) | FeNi metal (%) | Carbon-bearing phases | Hydrous phases | Reference |
|---|---|---|---|---|---|---|---|---|---|---|
| Enstatite | E | M | 0.01 ± 0.01 | 69.2 ± 6.5 | 10.5 ± 4.1 | -- | 20.3 ± 7.1 | x | x | Keil (1968) |
| Ordinary | H | S | 0.3 | 88.3 ± 1.2 | 3.7 ± 0.8 | 0.2 ± 0.1 | 7.9 ± 0.8 | | | Keil (1962) |
| | L | | 0.3 | 92.4 ± 1.2 | 4.1 ± 06 | 0.2 ± 0.1 | 3.3 ± 1.1 | | | |
| | LL | | 0.4 | 94.0 ± 1.0 | 4.1 ± 0.8 | 0.2 ± 0.1 | 1.8 ± 0.6 | | | |
| Carbonaceous | many | C (D, B, F, G, Q) | 0.4 | 93.1 ± 11.3 | 0.5 | 0.2 | 5.4 ± 7.1 | x | x | Bland et al. (2010) |
| Rumuruti | R | none identified | 0.7 | 92.3 ± 3.3 | 5.4 ± 3.0 | 0.8 ± 0.6 | 0.1 ± 0.1 | | trace | Kallemeyn et al. (1996) |
| Basaltic acondrites | Eucrites | V | 1.3 ± 0.5 | 99.8 ± 6.7 | 0.3 ± 0.2 | 0.9 ± 0.6 | -- | | | Mayne et al. (2009) |
| | Aubrites | E | 0.01 ± 0.01 | 98.0 ± 4.8 | 1.7 ± 2.7 | -- | 0.8 ± 1.4 | | | Mittlefehldt et al. (1998) |
| | Angrites | none identified | 0.6 ± 0.6 | 97.5 ± 0.7 | 1.2 ± 0.1 | 0.9 ± 0.3 | -- | | | Bischoff et al. (2000); Jambon et al. (2005) |
| Primitive achondrites | Acapulcoites & Lodranites | none identified | 0.11 ± 0.04 | 80.5 ± 9.3 | 4.5 ± 4.0 | 0.3 ± 0.4 | 10.5 ± 9.4 | x | x | Rubin (2007) |
| | Brachinites | none identified | 1-2 | 71.9 ± 40.3 | 3.1 ± 2.1 | -- | -- | | | Mittlefehldt et al. (1998) |
| | Ureilites | none identified | 0.3 ± 0.1 | 96.8 ± 4.3 | -- | -- | 1.3 ± 0.6 | x | | Singletary and Grove (2003) |
| Metal-rich | Mesosiderites | (S) (V) | 0.8 ± 0.9 | 49.6 ± 13.4 | 4.1 ± 3.4 | 0.8 ± 0.4 | 43.4 ± 18.9 | | | Mittlefehldt et al. (1998) |
| | Winonaites & Silicate-bearing irons | (M) | 0.1 | 42.5 ± 30.6 | 10.8 ± 13.7 | 2.1 ± 3.5 | 40.2 ± 28.4 | x | | Mittlefehldt et al. (1998) |
| | Pallasites | A | 0.2 | 65.0 ± 10.4 | 3.8 + 2.6 | 0.4 ± 0.5 | 30.8 ± 9.6 | | | Buseck (1977) |
| | Irons | M | -- | -- | <1 | -- | ~100 | x | | |
| Missions | Hayabusa | S | 0.4 | 87.2 | 11.3 | 1.2 | 0.3 | | | Nakamura et al. (2012) |
| | Stardust | -- | 0.5 ± 0.5 | 58.8 | 14.7 | | 5.9 | x | | Zolensky et al. (2006) |

Table 2. Threshold values and adjectival ratings of likelihood ratio (LR).

| LR value | Adjectival strength |
|---|---|
| $LR > 10^2$ | Decisive |
| $10^{1.5} < LR < 10^2$ | Very Strong |
| $10^1 < LR < 10^{1.5}$ | Strong |
| $10^{0.5} < LR < 10^1$ | Substantial |
| $10^0 < LR < 10^{0.5}$ | Weak |
| $LR < 10^0$ | Very Weak |

Table 3. Test case meteorite phase abundances, most likely model classification, and number of particles *n* in the sample required to yield the most likely class (and subclass) in 90% of trials. In general, tens to hundreds of particles are sufficient to classify a sample using only these parameters, but in some cases additional information would be useful.

| Meteorite | Class | Silicate Fe/Mg | Silicate | Sulfides + phosphates | Oxides | FeNi metal | Model class (subclass) | *n* (90%) |
|---|---|---|---|---|---|---|---|---|
| Bluff | L5 | 0.25 | 92.2 | 4.2 | 0.36 | 3.2 | Ordinary chondrite (L) | 100 (200) |
| Bath | H4 | 0.30 | 88.8 | 3.6 | 0.09 | 7.5 | Ordinary chondrite (H) | 100 (200) |
| Abee | Enstatite | 0.02 | 58.8 | 17.4 | 0.00 | 23.8 | Enstatite | 50 |
| Juvinas | Eucrite | 1.57 | 99.1 | 0.2 | 0.70 | 0.0 | Eucrite | 20 |
| ALH A77081 | Acapulcoite | 0.12 | 88.2 | 5.2 | 1.30 | 4.4 | Ordinary chondrite (L) | 100 (700) |
| Dhofar 1222 | Acapulcoite | 0.07 | 73.8 | 6.4 | 0.30 | 16.5 | Enstatite | 200 |
| Hayabusa | LL5 | | | | | | Ordinary chondrite (LL) | 100 (200) |

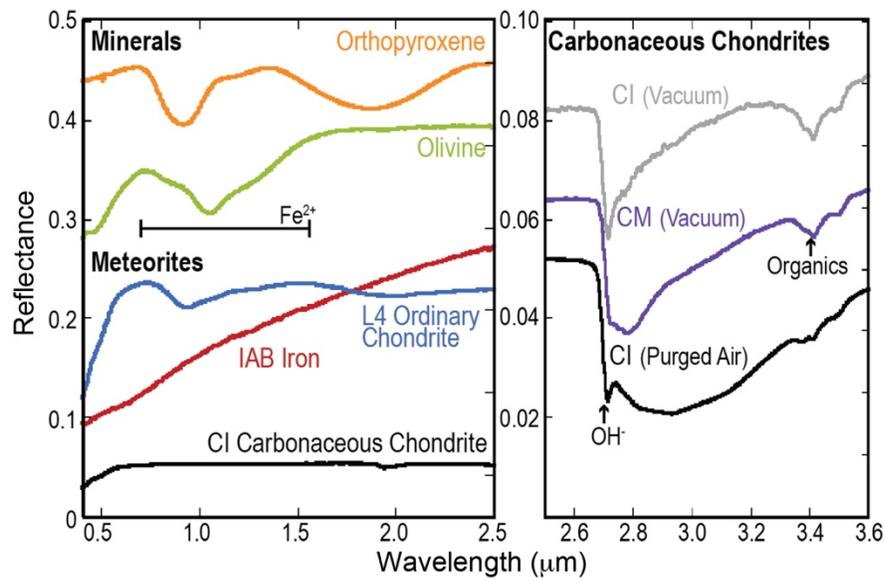

Figure 1. Example of silicate, carbonate, organic, and accessory mineral spectral features in the near IR region. Data sourced from (McAdam et al., 2015; Takir et al., 2013) and the RELAB database.

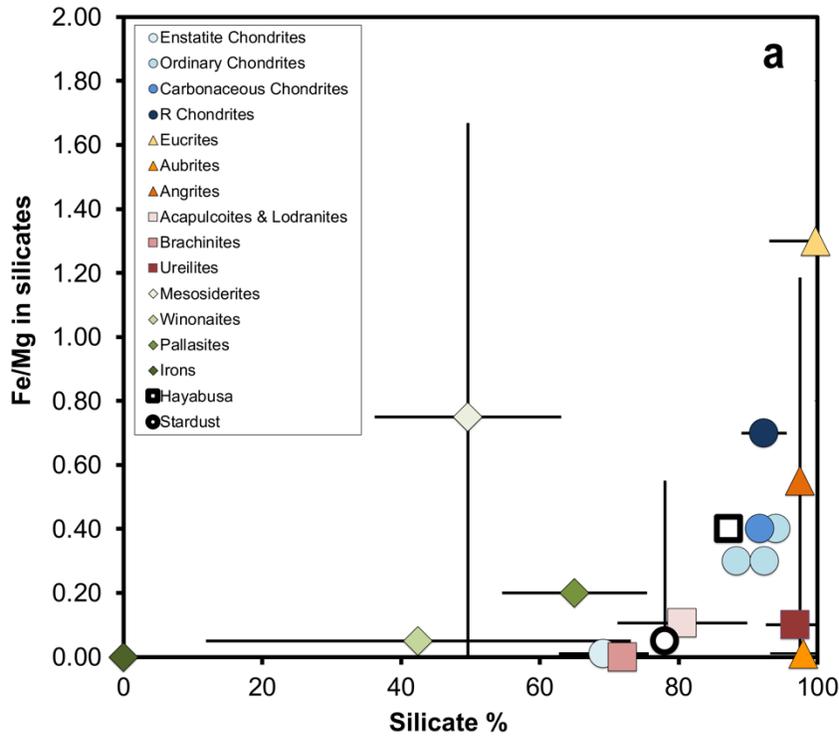
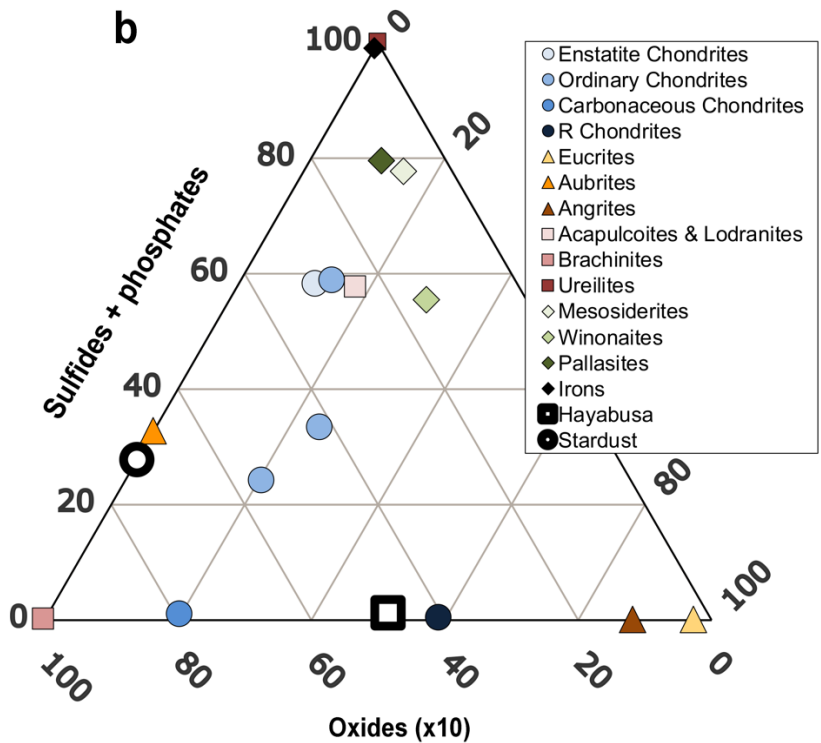

Figure 2. Meteorite groups plotted as simple combinations of a) silicate abundance and Fe/Mg ratio in the silicate fraction (unequilibrated meteorites have large ranges in Fe/Mg ratios); and b) sulfides + phosphates, oxides, and FeNi metal.

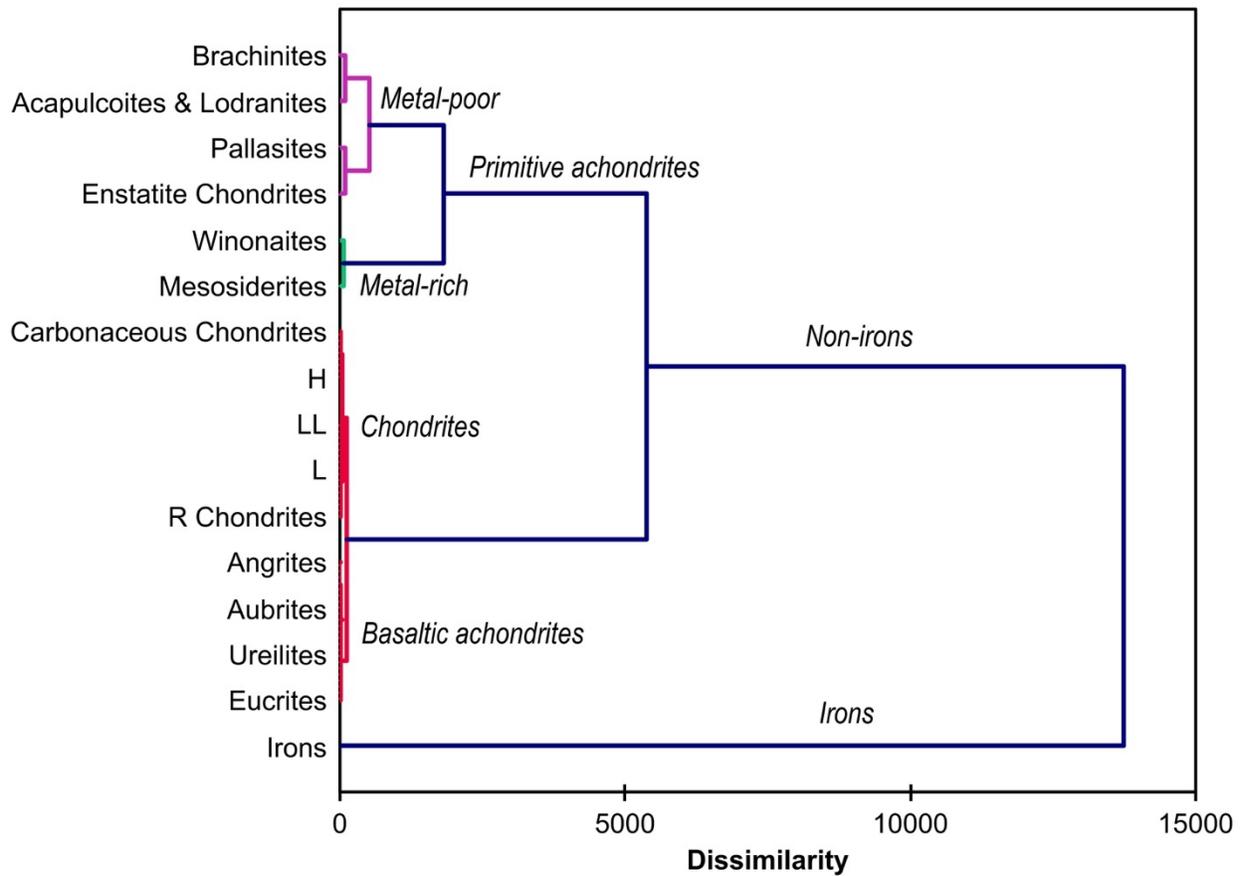

Figure 3. Dendrogram showing the results of Agglomerative Hierarchical Clustering of meteorite classes in Table 1 using only the average abundance of four mineral phases (silicate, sulfides + phosphates, oxides, and Fe-Ni metal). Each merge is represented by a horizontal line. The distance along the x-axis represents the similarity of the clusters that were merged. In general, multidimensional fitting using these parameters distinguishes major groups of meteorites from each other.

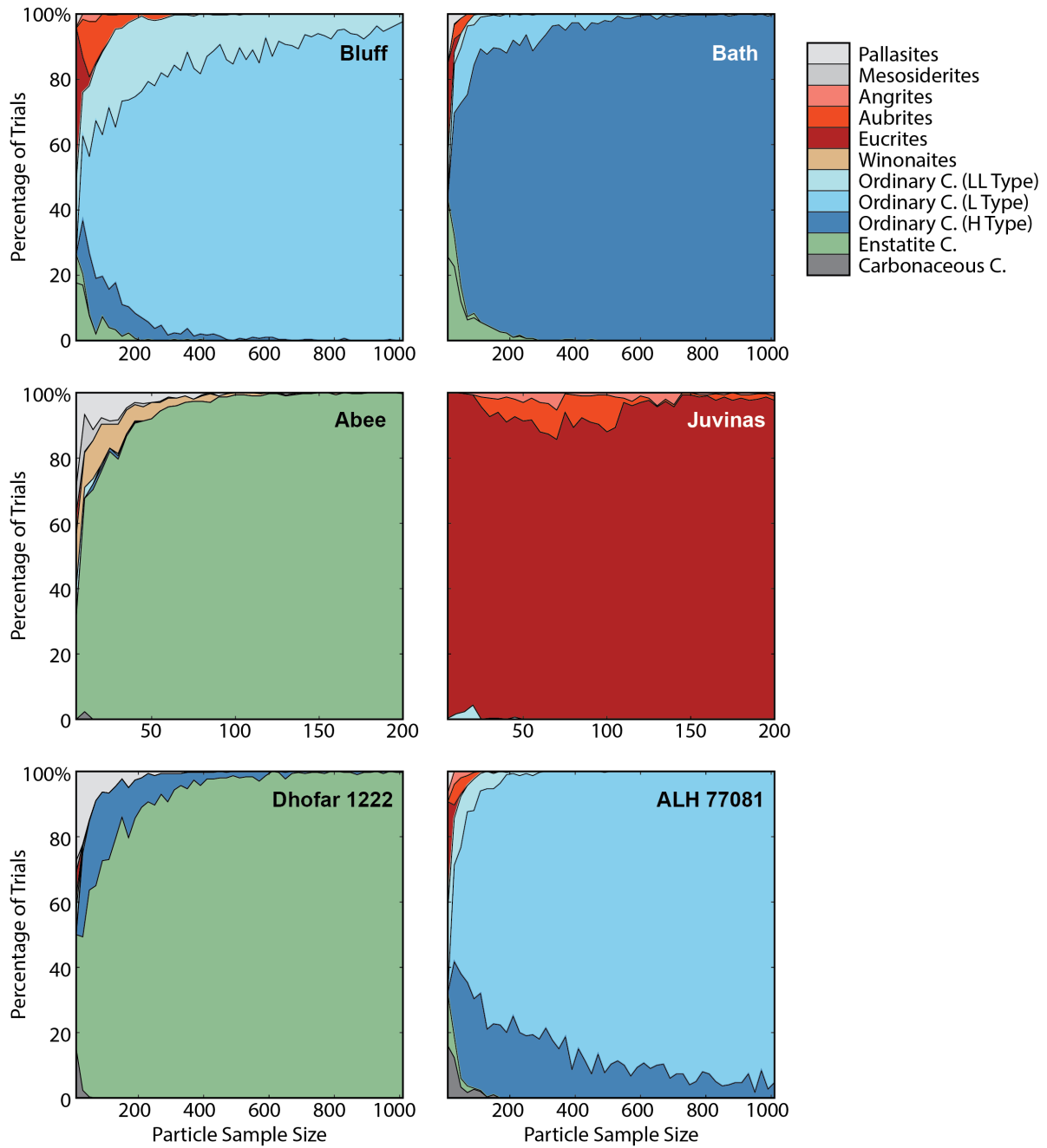

Figure 4. Classification results for model datasets created from known meteorites. Bluff and Bath are linked to ordinary chondrites in about 100 samples, and to their subclass in about 200 samples. Abee and Juvinas are shown to be an enstatite chondrite and a eucrite using only 10's of samples. However, metal-poor acapulcoites Dhofar 1222 and ALH 77081 are confused with other classes.

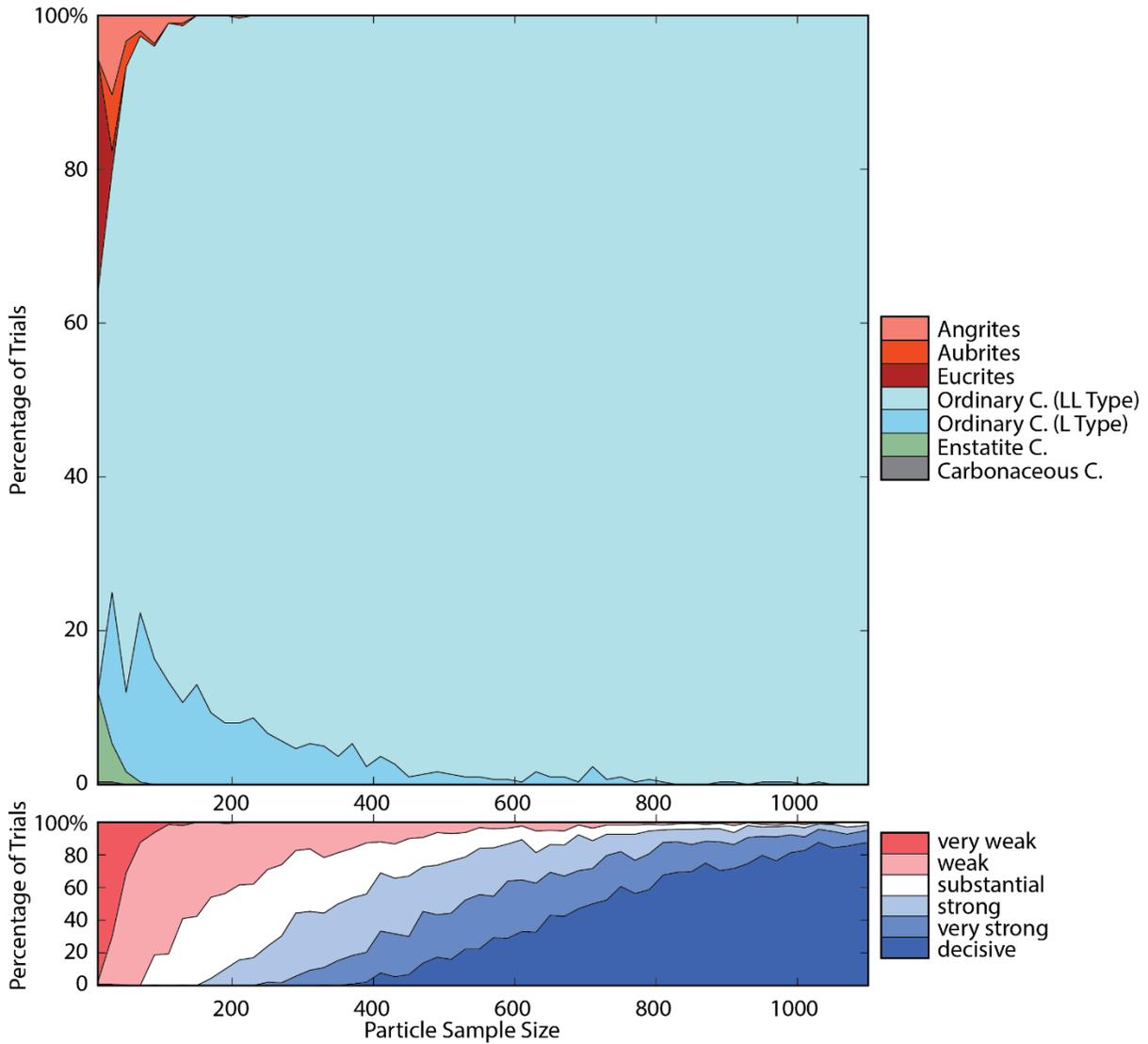

Figure 5. (top) Selected meteorite classes for individual trials (y axis) of sample sets with *n* from 20-1100 (x axis) selected from the Hayabusa returned sample (Table 1). As *n* increases, the number of trials corresponding to chondrites and specifically LL-chondrites, increases. (Bottom) the evidential strength of the selected meteorite class for each individual trial, no matter the guessed type. For small numbers of particles, the evidence to suggest that the sample matches one meteorite class over another is generally weak, but increases with increasing *n*.

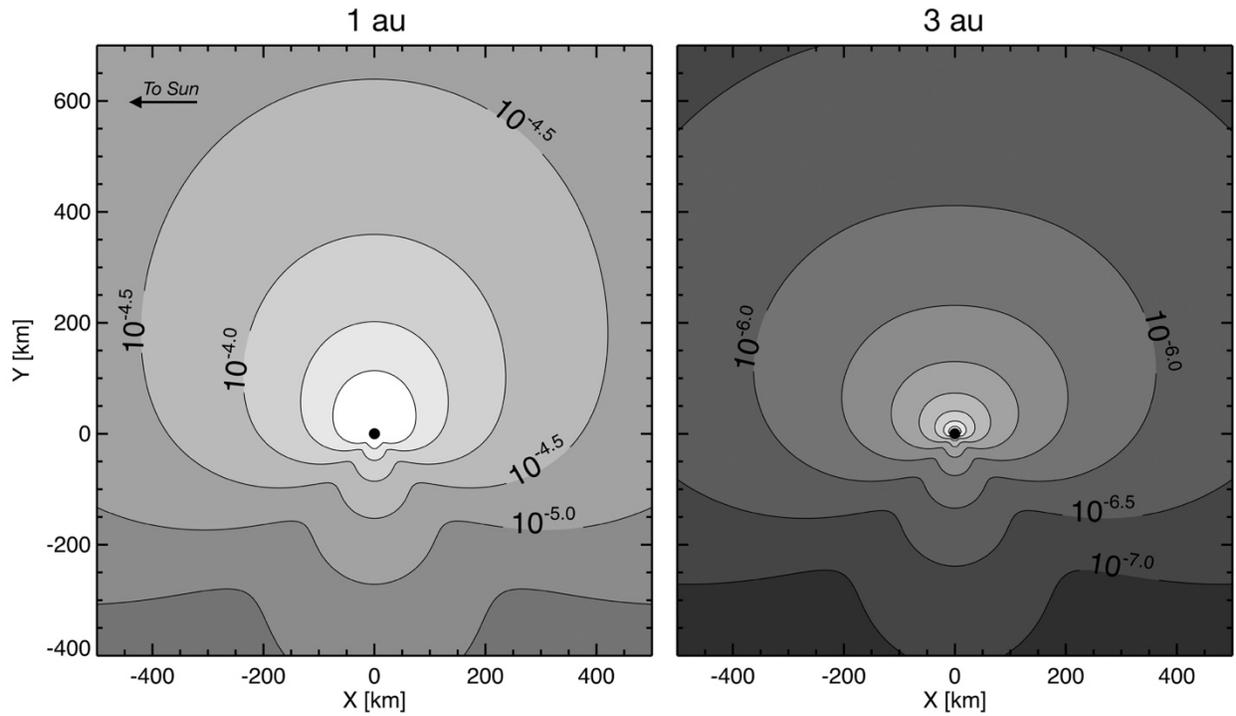

Figure 6. Comparison of the ejecta cloud structure, in particles/m$^3$, for a 10 km radius body at 1 au and 3 au for grain radii a > 50 nm. The reference frame shown has the Sun fixed in the –x direction and the apex direction in the +y direction. Due to the expected larger decrease in retrograde HTC particles compared to JFCs producing the HE/AH ejecta, ejecta from the apex source is reduced at 3 au producing a flattened density distribution.

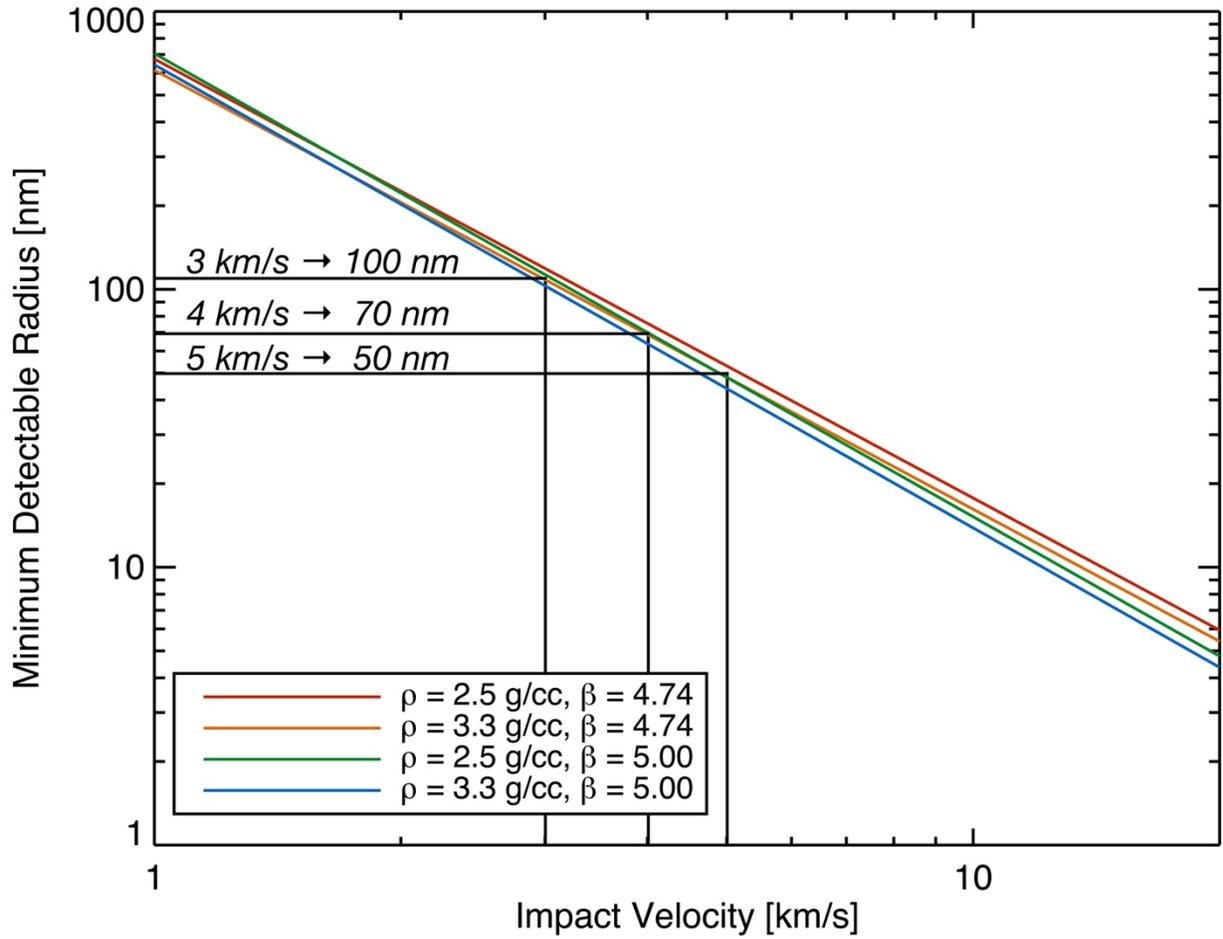
Figure 7. Minimum detectable dust grain radius as a function of impact velocity.

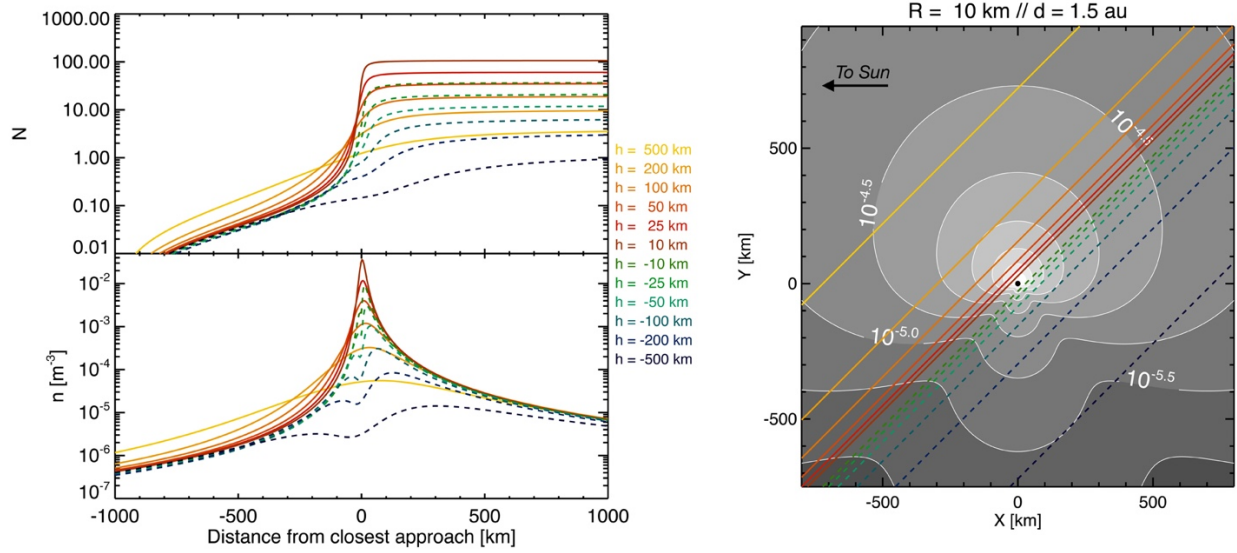

Figure 8. Total number of impacts and predicted dust density (a > 50 nm) for a 10 km radius body at 1.5 au. Contours are logarithmically spaced in units of m$^{-3}$.

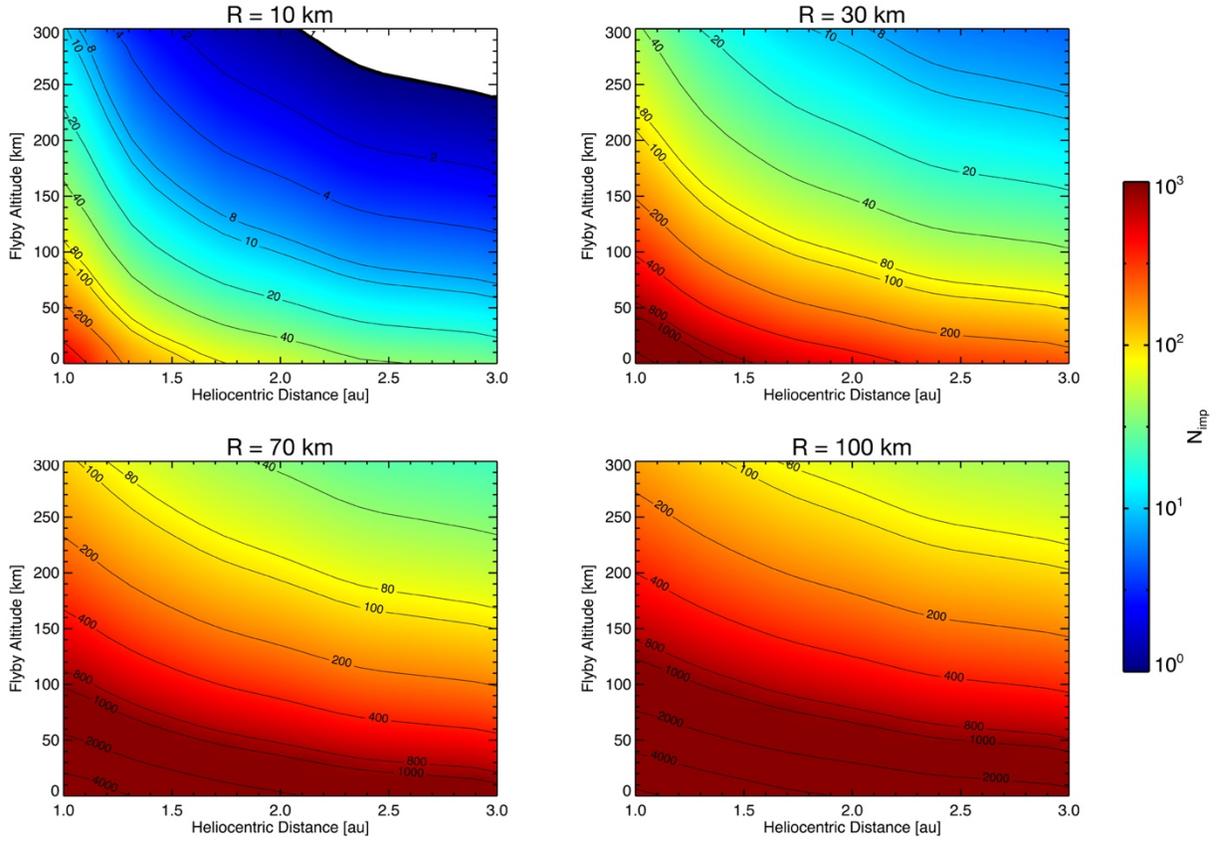

Figure 9. Total impact count predictions for grain size a > 50 nm as a function of heliocentric distance and flyby altitude for R = 10, 30, 70, and 100 km.

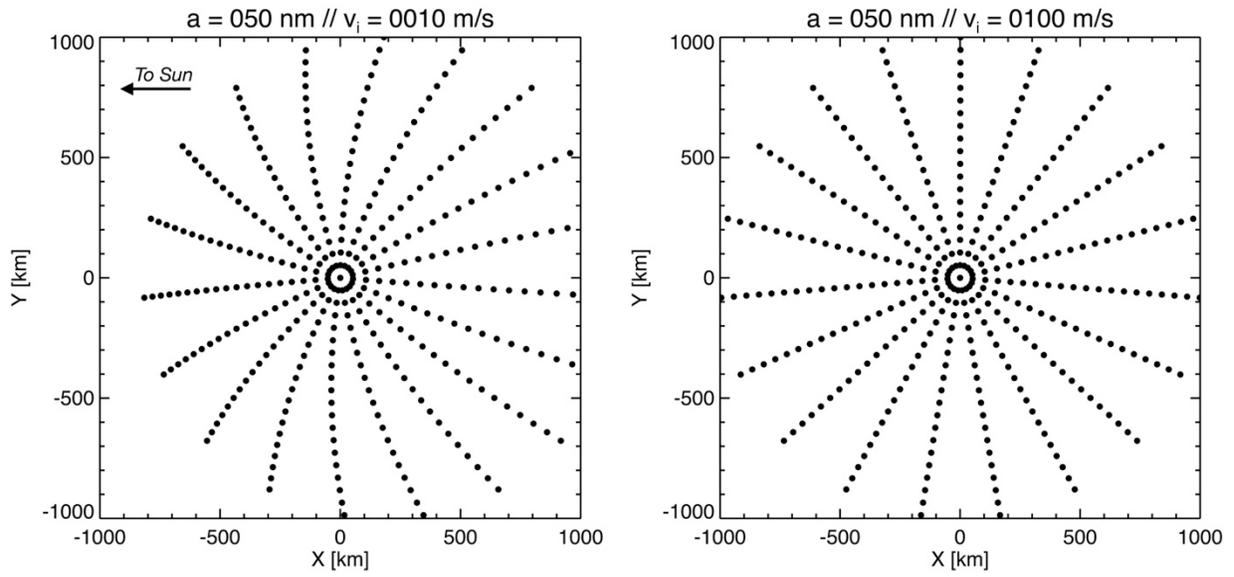

Figure 10. Simple simulation showing the effects of radiation pressure on ejecta dynamics, with the dots representing positions of grains launched from the surface at different times. Left and right are given for initial speeds of 10 m/s and 100 m/s respectively.